# Adsorption – desorption processes on mesoscopic pores connected to microscopic pores of complex geometry using the Ising model

M. A. Balderas Altamirano[*1, 3], S. Cordero[1], G. Román[2] and A. Gama Goicochea[3]

*(1) Departamento de Química, (2) Departamento de Ingeniería Eléctrica, Universidad Autónoma Metropolitana, Unidad Iztapalapa, Av. San Rafael Atlixco No. 186, México, D. F. 09340; (3) Instituto de Física, Universidad Autónoma de San Luis Potosí, Av. Álvaro Obregón 64, 78000 San Luis Potosí, México*

[*] Corresponding author; e-mail: nyfg@xanum.uam.mx

**ABSTRACT:** In this work we report studies of nitrogen adsorption and desorption onto solid surfaces using computer simulations of the three dimensional Ising model, for systems with complex porous structures at the mesoscopic and microscopic levels. A hysteresis cycle between the adsorption and desorption processes appears and we find that its characteristics are dependent on the geometry of the pore and on the strength of the surface – fluid interaction. We obtained also an average adsorption isotherm, which represents a combination of differently shaped pores, and shows robust jumps at certain values of the chemical potential as a consequence of the structures of the pores. Lastly, we compare our results with experimental data and also report the filling process of microscopic pores connected with mesopores. It is argued that these predictions are useful for researchers working on the enhanced recovery of oil and for the design of new nanomaterials, among others.

# I. INTRODUCTION

The contemporary importance of the study of adsorption has its origins in applications such as the drying of gases and liquids by activated alumina, production of nitrogen from air by means of carbon molecular sieves, biocompatibility in materials, water treatment, environmental applications, and adsorption heat pumps, to mention but a few (Dabrowski, 2001). When a material is filled with vapor through an increase in the pressure or the chemical potential, an adsorption isotherm can be obtained; by reversing the process one gets a desorption curve and a hysteresis cycle is found. The mechanisms leading to and having impact on adsorption hysteresis are still not yet completely understood, which warrants the need for further studies. Three models are generally used for the interpretation of hysteresis in mesoporous materials: independent pores (Sarkisov *et al*. 2001), pore network (Gil Cruz *et al*. 2010), and disordered pores (Lowell *et al*. 2004). The first one considers that the hysteresis cycle arises from the formation of the liquid-vapor interface, also called the meniscus formation, and through the condensation process a cylindrical interface is formed, while a spherical phase forms during the evaporation process (Rasmussen *et al*. 2010, Monson, 2009). The pore network is made up of the interconnectivity of pores; a sharp step on the desorption isotherm is usually understood as a sign of interconnection of the pores. A pore connected to the external vapor phase via a smaller pore acts as a neck (often referred to as an ink bottle pore). Complex disordered pores require knowledge of the porous network in addition to the complex geometry of the pore for a thorough understanding (Thommes *et al*. 2006, Nguyen *et al*. 2013). The relatively recent development of porous materials with specific structures, such as those called SBA 15 (Zhao *et al*. 1996 and 1998, Yanagisawa *et al*. 1990) and the MCM 41 (Beck *et al*. 1992) has helped advance the understanding of the hysteresis cycle in porous materials, because they are engineered with large surface areas and pore diameter in the range 2 - 30 nm. These developments have advanced the path to new applications in areas as diverse as drug delivery and carbon adsorption (Ge *et al*. 2011, Attili *et al*. 2013). A comprehensive review of the experimental aspects of this material can find in Yue *et al*. (2008).

High performance computing technologies have opened up the way to the modeling of large adsorption systems. These complemented efforts (experimental, theoretical and computational) have greatly advanced our understandings of the phenomenon of sorption on complex structures. One of the models most frequently used is density functional theory; a recent review has been published by Landers *et al*. (2013). The development of models like the so – called dual site bond model (DSBM, Mayagoitia *et al*. 1985), which represents a complex porous solid material, have proved to be fruitful also ( Mayagoitia *et al*. 1988, Rojas *et al*. 2002, Gil Cruz *et al*. 2010, Cordero *et al*. 2005). Yet another model that has shown to be successful for the study of the adsorption/desorption phenomenon is the well-known three dimensional (3D) Ising model (Ising, 1925). Several authors have applied the 3D Ising model to the study of adsorption on complex structures, such as Edison *et al*. (2009) who considered a carbon slit and simulated the hysteresis cycle adding an anisotropic point in the system. Naumov *et al*. (2009) added roughness to a cylindrical pore, while Pasinetti *et al*. (2005) modeled the adsorption in a one dimensional channel.

The purpose of this work is to report numerical simulations that yield an improved understanding of the adsorption and desorption processes in complex pore distributions through the variation of the interactions between the surface and the fluid, as well as the construction of mesoscopic structures. To accomplish this task we have constructed pores with various geometrical shapes and sizes and have obtained adsorption/desorption isotherms as functions of the chemical potential, strength of the surface – gas interaction, and of the gas – gas interaction, using Monte Carlo simulations. This article is organized as follows: in Section II we present the details of the model and the technique used to solve it, as well as the geometrical characteristics of all the pores simulated. Section III is devoted to the presentation and discussion of our results. Firstly we test our program's ability to reach equilibrium quickly, then vary the fluid – solid interaction and compare with experimental data. In the second part we present the structures we have modeled and obtain an average to reproduce the complex adsorption and desorption processes that occur in actual pores. In the third section we model a structure with radii in the range from microscopic to mesoscopic scales, and show how the adsorption and desorption processes are modifies by this complex network. Finally, the conclusions can be found in Section IV.

**II. MODELS AND METHODS**

We use here the 3D Ising model on a bcc network (Newman *et al*. 1999), with sites representing either an empty place or a gas molecule adsorbed on the surface, interacting with their closest neighbors. An initial site can take only two values: 0 for an empty site and 1 for one filled with a gas molecule. The Hamiltonian used in this work is (Woo, *et al*. 2004):

$$H = -J\sum_{i,j} n_i n_j t_i t_j - \mu \sum_i n_i t_i - yJ \sum_{i,j} n_i t_i (1-t_j) + n_j t_j (1-t_i) \tag{1}$$

where $n_i = 0, 1$ is the fluid occupation variable of the lattice gas, $t_i$ is the variable representing the solid matrix configuration, equal to either 1 or 0, depending on whether the site is available for occupancy by the fluid, or blocked by the solid, respectively. The summations run over all of the nearest neighbor site pairs; the constant $J$ sets the strength of the fluid – fluid interaction, and $\mu$ is the chemical potential. The third term in eq. 1 accounts for the additional attractive interactions between the adsorbed particles and the solid surfaces, whose strength can be varied by adjusting the *y* parameter. The methodology followed for the implementation of the 3D Ising model was taken from Newman *et al*. 1999.

We have modeled three pore structures, shown in Fig. 1. The structure seen in Fig. 1(a) is a cylindrical pore with sinusoidal oscillations in the *z* – axis. There are two radii, R1 and R2, with R1 being larger than R2, defined as the distance from crest to crest between both walls, and R2 is the valley to valley distance. The pore in Fig. 1(b) is composed of a cylinder interconnected with a sphere of radius R1, with the cylinder having a radius equal to R2. The structure shown in Fig. 1(c) is an interconnected parallelepiped box. In structures (b) and (c) there are 5 units (each unit is either a sphere, a cylinder or a sinusoidal region) with characteristic length equal to R1, and 6 units with R2 interconnected with a vapor phase. The cell size has several values, which can be found in Table I. At least two layers of solid molecules represent the pore structure. We used periodic boundary conditions on the three axes. Throughout this work we use reduced units, namely $U^*=U/k_BT_c$, $T^* =T/T_c$ and $\mu^*=\mu/k_BT_c$, where $U$ is the potential energy of the system, $k_B$ is Boltzmann´s constant, $T_c$ is the critical temperature of nitrogen, equal to 127 K. All distances are reduced with the effective diameter of the nitrogen molecule, $\sigma = 0.3$ Å (Woo et al. 2004).

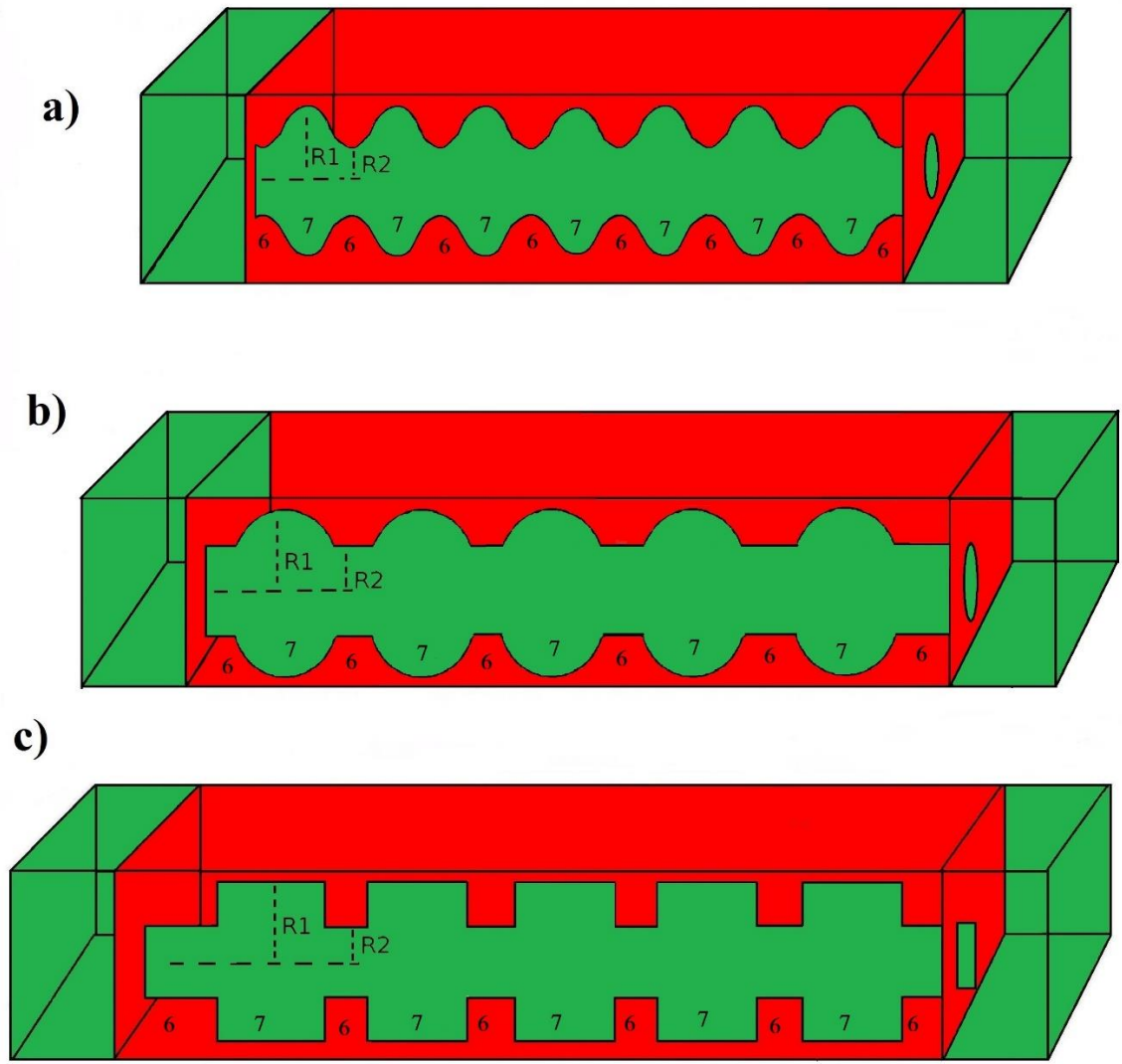

**Figure 1**. Our model pore structures. (a) Sinusoidal pore. (b) Cylinder – sphere pore. (c) Parallelepiped pore. R1 and R2 are the radii or the half of the length of the cavity in the pore, which for simplicity are called radii even when there is no curvature, as in (c). We constructed also cylindrical structures that have R1=R2 (not shown), which are used for comparison with those shown in this figure, see Table I. The numerical sequence in each pore represents a case with R1=7 and R2=6.

Once each cell is constructed the grand canonical ensemble (fixed $\mu$, volume $V$, and $T$) is applied. The simulations for each structure reported in this work were run for at least $10^4$ Monte Carlo steps (MCS); in each MCS there were $N$ (number of molecules) iterations to include a vapor molecule into the adsorption system, leading effectively to $N \times 10^4$ MCS. The first $N \times 10^3$ MCS were used to equilibrate the simulation, and the rest were used to calculate averages. To optimize the performance of the simulations, each configuration was started from the final configuration of the previous simulation. The calculations started with the smallest chemical potential necessary to have an almost empty system, then we increased the chemical potential at fixed increments of 0.01 units until finally the system was filled with fluid. Then, the chemical potential was reduced gradually to empty the structure. The temperature modeled in this simulation was set to $T^* = 0.6$, which is equivalent to the value at which many adsorption experiments are carried out ($T = 76$ K). The values of $y$ were chosen as 2.0, 1.5, 1.2 and 0.9, while $J = 0.7447$ which is the nitrogen interaction constant (Ravikovitch *et al*. 1997, and 2002). The length of each structure depends on the volume of the cylindrical structure taken as reference, which is a structure with R1=R2. The simulations stop when the main structure is full of absorptive, and then it is emptied so that a desorption curve can be obtained also.

**Table I**. Composition R1 and R2 for each structure simulated. R1 and R2, represent the half of the length, or the radii in the porous structures, see Fig. 1. R1 is always the larger radius or length in the structures, and R2 is the smaller one. When R1=R2, the structure is a perfect cylindrical pore. Each pair of R1 and R2 combinations with the a), b) or c) structures (Fig. 1), represents a simulation carried out in this work. The values of the radii are reported in reduced units, $\sigma$.

| R1 | R2 | R1 | R2 | R1 | R2 |
|----|----|----|----|----|----|
| 7  | 7  |    |    |    |    |
| 7  | 6  | 6  | 6  |    |    |
| 7  | 5  | 6  | 5  | 5  | 5  |
| 7  | 4  | 6  | 4  | 5  | 4  |
| 7  | 3  | 6  | 3  | 5  | 3  |
| 7  | 2  | 6  | 2  | 5  | 2  |
| 7  | 1  | 6  | 1  | 5  | 1  |

Once all simulations were performed, we summed up all isotherms and divided by the total volume, thereby creating an average isotherm to represent them all. We did this for each group of simulations that we present in Table I.

## III. RESULTS AND DISCUSIONS

*I. Performance of the 3D Ising model to predict adsorption – desorption processes*

In the first part of this work, we carried out a set of simulations to test the methodology. In Fig. 2, we show the average particle number <N> in a cylindrical pore with R1=R2=7$\sigma$ as a function of the MCS. As one can see in Fig. 2, the oscillation are rather strong when the MCS are relatively small, therefore we chose $10^4$ x (Network size) x MCS, to achieve better performance. To further improve the performance, we used the last configuration from the previous block to begin the next one, thereby avoiding the need to start from the initial configuration. This process leads to a more efficient way to reach the equilibrium of the system.

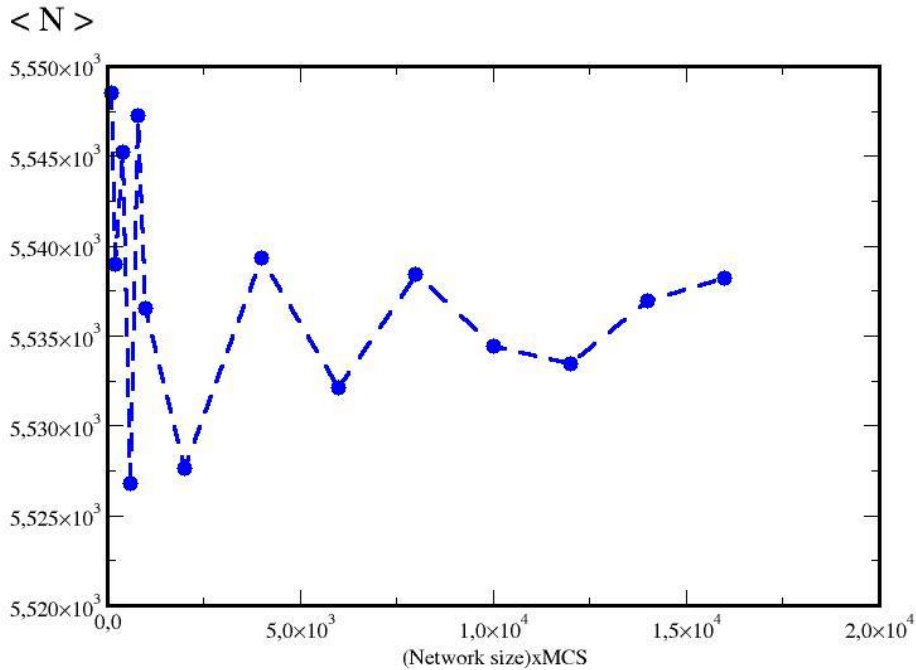

**Figure 2**. Average particle number <N> vs number of Monte Carlo steps, for $\mu^*$=-3.1, at constant temperature $T^*$ = 0.6, with $J$ = 0.7447 and $y$ = 1.5, with R1 = R2 = 7$\sigma$. Lines are only guides for the eye. Network size means the number of sites in the network, which is given by the product $L_x \times L_y \times L_z$, where $L_x$, $L_y$ and $L_z$, are the length of the simulation box in the $x$ -, $y$ -, and $z$ − axes, respectively.

To explore the influence of the surface – fluid interaction, *y*, we carried out simulations of adsorption/desorption in a perfect cylindrical pore with radius equal to 15 σ and varying values of *y*, namely *y* = 0.9, 1.2 and 1.5, and the results are presented in Fig. 3. At *y* = 0.9 the isotherm shows very poor adsorption until the reduced pressure (P/P$^o$) is close to 0.78, then the vapor condenses into a liquid. Clearly, the interaction between the pore and the gas molecules is too weak, at least within the scope of the model of eq. (1), which takes into account only nearest – neighbor interactions. For *y* = 1.2 and *y* = 1.5 the isotherms form multiple layers from P/P$^o$ = 0.4 up to 0.78, as indicated by the adsorption line before the condensation, at P/P$^o$ = 0.78. For *y* = 1.5, the isotherm shows adsorption of more vapor molecules before the condensation process begins, which occurs at about P/P$^o$ = 0.45. This trend is to be expected since by increasing the *y* value a more attractive wall is obtained, therefore the vapor molecules can be trapped more easily on the surface of the pore until they condense into a liquid.

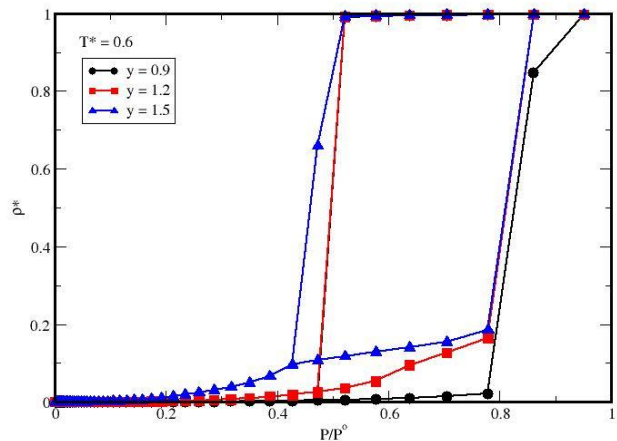

**Figure 3**. Effect of the *y* parameter (see eq. (1)) on the adsorption over perfect cylindrical structures of radius equal to 15σ. Filled circles correspond to *y* = 0.9, filled squares are data corresponding to *y* = 1.2 and the line with triangles is for *y* = 1.5. In all cases *T*\*=0.6 and *J* = 0.7447, see eq. (1). Lines are guides to the eye, and the axes are shown in reduced units.

We test the method by comparing with experimental results; we constructed a cylindrical pore with radius equal to 2.1 nm to compare with recently published work reported by Ojeda *et al.* 2003. To do so we fixed for all cases the temperature at 0.6 and the *y* parameter at 2.0, while *J* = 0.7447. In Fig. 4, we show the isotherm obtained for this case. The line with circles represents our data while the line with squares represents the experimental results of Ojeda *et al.* 2003. In this figure a kink at a reduced pressure of approximately 0.08 appears, which signals the formation of a monolayer; this is to be compared with the results of Ojeda *et al.* 2003, who found that a monolayer forms at P/P$^0$ = 0.1. It is remarkable that similar trends between those experiments and our calculations are obtained using a relatively simple model, which nevertheless is capable of capturing meaningful thermodynamic insights into the nature of the hysteresis cycle. It is important to recall that eq. (1) incorporates the only interactions that are relevant in a minimal model such as ours, namely the fluid – fluid and the surface – fluid interactions.

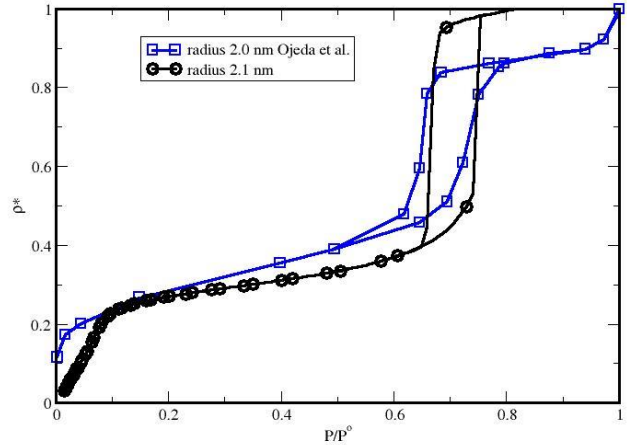

**Figure 4**. Adsorption-desorption isotherms of gas on a perfect cylinder, with $y = 2.0$, $J = 0.7447$ and $T^* = 0.6$. The filled circles are results from the present work while and the squares are data taken from Ojeda *et al*. 2003. The lines are guides to the eye. The $x$ – axis represents the pressure (P) of the pore relative to the saturation pressure ($P^0$), which is how the chemical potential in the pore is modified.

*II Understanding the filling and emptying processes: adsorption – desorption in pore structures.*

To understand the filling and emptying processes, we choose a structure like the one shown in Fig. 1(c), with radii R1 and R2 whose values are indicated in the inset of Fig. 5. The process of adsorption and desorption can be understood as follows. The isotherm was obtained from our simulations, choosing R1=7, 6 and 5σ, and R2= 3σ. The structure shown represents micropores connected with a mesopore material, according to the IUPAC (Sing *et al*. 1985). Following its recommendations we can identify in Fig. 5 the micropore, mesopore and macropore zones in the isotherm. According to the IUPAC definitions, pores with at least 3σ of radius are micropores; mesopore zones are defined as those with radii of up to 83 σ, and the macropore zone involves pore whose radii are larger than 83 σ. To better identify each zone we have labelled them with Greek letters. At small values of the pressure there is the solid pore sample and as we add gas molecules they start interacting with the solid, and begin to form a monolayer; this process is completed when P/P$^o$ is about 0.3. If one increases the pressure or the chemical potential, more molecules are added to the system and a slight plateau appears from 0.3 to 0.4, which means the filling process is adding more molecules to form a multilayer. In Fig. 5, in the region labeled 3α) a kink appears, which is a consequence of the filling of the pore with radius 3σ, 0.42 P/Pº, see also Fig 6α).

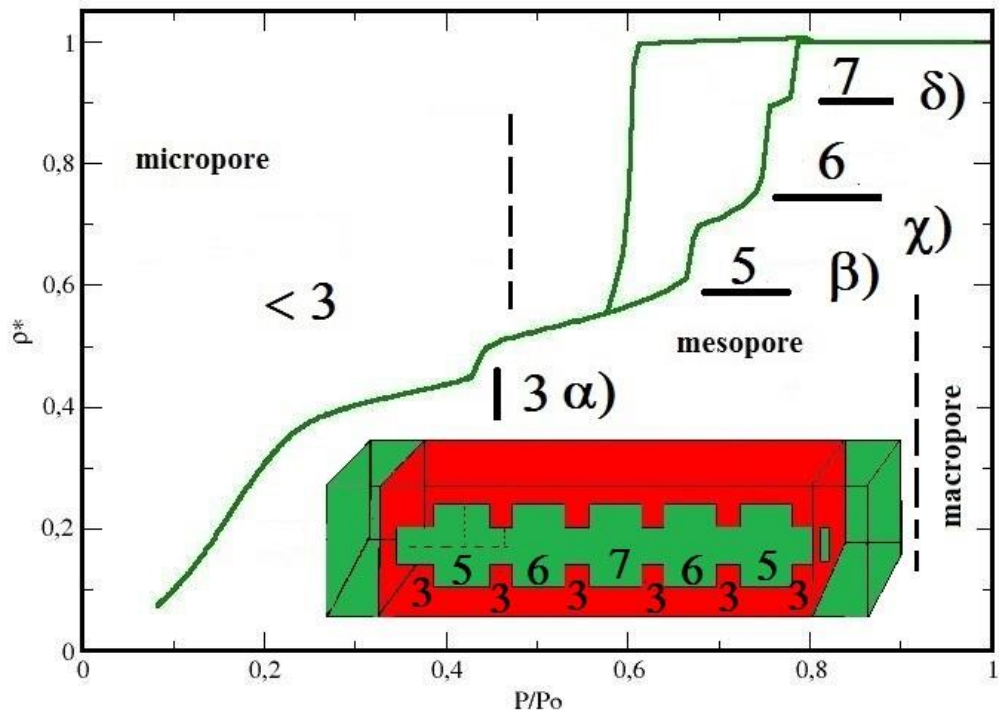

**Figure 5**. Isotherm for a complex pore network, which was built like the structure in Figure 1(c). The $x$ – axis is P/Pº and the $y$ – axis is particle density in the system. The inset represent the pore structure used for this isotherm. The Greek letters represent the zones where pores are completely filled with liquid at the radius indicated (numbers). The division of sections of this isotherm into zones called micropore, mesopore and macropore follows from the recommendation of the IUPAC, see Sing *et al*. (1985).

In Fig. 6 we show four graphs, each represents the average number of particles <N> ($y$ – axis, graph Fig 6) present in the pore as a function of the length (simulation box, $z$ – axis) of the pore ($x$ – axis, graph Fig. 6) which is the one seen in the inset of Fig. 5. The Greek letters correspond to the zones labelled also with Greek letters in Fig. 5, with their corresponding radii given in units of σ. The purpose of Fig. 6 is to show how the different pores in the structure shown in the inset of Fig. 5 get filled up at different relative pressure (or chemical potential). The black line represents the number of particles at the relative pressure, P/Pº, before the condensation, and the red one after the condensation. Fig. 6α), shows the filling of the micropore, or pore with radius equal 3σ. The filling of this pore does not occur in one step, it needs a range of pressures, from 0.43 to 0.61 P/Pº to fill up. It should be remarked that at, although there are six identical pores with radius equal to 3σ, it takes an additional increment of the relative pressure, up to 0.61 P/Pº, to fill the other four pores of radius equal to 3σ. When the pressure is further increased, more gas molecules are added into the structure, and a kink appears at P/Pº = 0.67 in the zone labelled β) in Fig. 5, which corresponds to the filling of pore with radius of 5σ, in Fig. 6 β), which shows the number of particles before the condensation takes place, at pressure is 0.66 P/Pº (black line), and just after, at 0.67 P/Pº (red line). Then, a slight plateau appears when the relative pressure goes from 0.67 to 0.72, see the adsorption isotherm in Fig. 5, just before the zone labelled 6χ). When we increase the pressure a new kink at 0.76 P/Pº appears in Fig. 5, which corresponds to the complete filling of the pores with radius equal to 5σ in Fig. 6 χ), and the incipient filling of the pores with radius equal to 6σ (black line) at relative pressure of 0.75 P/Pº, followed by the complete filling of these pores (with radius of 6σ) at 0.76 P/Pº (red line). Finally, the filling of the largest pore, with radius of 7σ, occurs at 0.79 P/Pº, see Fig 6δ), which corresponds to the step before the relative pressure is 0.78 and after it is equal 0.79 P/Pº in the zone labelled 7δ) in Fig. 5., where the kink in the adsorption isotherm signals the condensation of the pore of radius 7σ. The reverse process, desorption, does not trace back the adsorption line, as expected, until all the pores with radii larger than 3σ have been emptied.

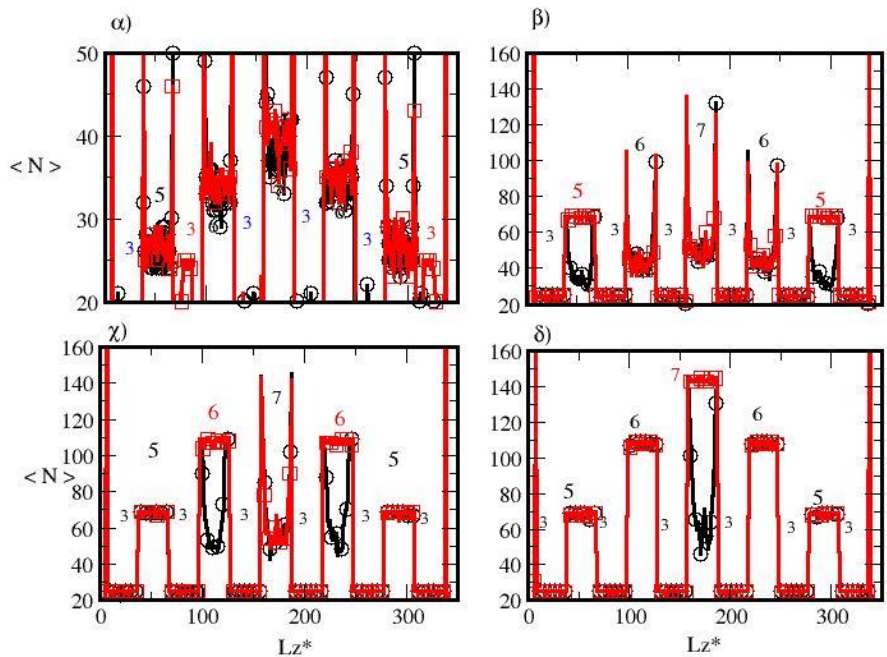

**Figure 6**. Filling process, illustrated by the <N> profile, for the structure shown in Fig 5. The line with circles represents the P/P° before and after condensation takes place, at relative pressures equal to α) 0.43, 0.43; β) 0.66, 0.67, χ) 0.75, 0.76 and δ) 0.78, 0.79, respectively. In the case shown in the panel labelled α), the chemical potential between the two cases shown is actually different, although in terms of the relative pressures such change appears to be too small. The black line is the filling of the pores before condensation occurs, while the red line represents the particle number just after condensation takes place.

*III Simulating complex porous materials*

The sorption graphs shown in Fig. 7 represent the isotherms obtained from simulations of the adsorption/desorption processes for only the three mesopore structures shown in Fig. 1, and defined in Table I, at fixed radii, with $T^*=0.6$, $y=2.0$ and $J=0.7447$. We choose only three structures to show how qualitatively different isotherms are obtained from different structures. In Fig. 7a) we present the sinusoidal structure (see Fig. 1a), with R1=7 and R2=6. This isotherm shows the filling and emptying processes, which are reversible up to $\mu^*= -3.25$, see Fig.7a), then there is condensation at $\mu^* = -3.17$ approximately; the reverse process produces the hysteresis cycle and evaporation occurs at $\mu^*= -3.25$. Fig. 7b) is the isotherm obtained for the structure seen in Fig. 1b), an interconnection of spherical and cylindrical pores with R1=7 and R2=4. The filling and emptying processes are different from the isotherm shown in Fig. 7a). One can see that the filling and emptying processes are reversible up to $\mu^* = -3.5$. Adding more gas molecules produces a kink at $\mu^*= -3.4$, which corresponds to the filling process of the smaller radius, in this case R2=4 (as we have already seen, in Fig. 5 and Fig. 6). Increasing the pressure raises the number of gas molecules into the structure, and a plateau and then another kink appear at -3.4 $\mu^*$, this filling process belongs to the second radius R1, filled with gas molecules. Since we have two different radii, the structure can develop at least two kinks in the isotherm. This is true for all the cases, since we have variables like length, temperature, and other thermodynamic interactions ($y$, $J$ and $\mu$ in equation (1)) set equal for the all the structures whose isotherms are shown in Fig. 7. Yet, their hysteresis cycles turn out to be different, because the geometrical shape of the pores plays a crucial role, even when the size of the structures and the thermodynamic conditions of their filling and emptying are the same.

In Fig. 7(c), we present the isotherm of the rectangular interconnections, shown in Fig. 1(c). One finds a slightly larger hysteresis cycle than those seen in Figs. 7(a) and 7(b). Fig. 7(c) shows additionally a kink at $\mu^*=-3.65$, representing the filling of the R2 radius, as we explained in Figs. 5 and 6, for a specific structure, see inset in Fig. 5. Adding more gas molecules leads to a kink at $\mu^*=-3.22$, which belong to the filling of the pore with radius equal to 7σ. The differences seen in the three curves shown in Fig. 7 are attributed to the geometrical differences of the various cases shown in Fig. 1 and table 1, because the interactions in eq. (1) are kept the same.

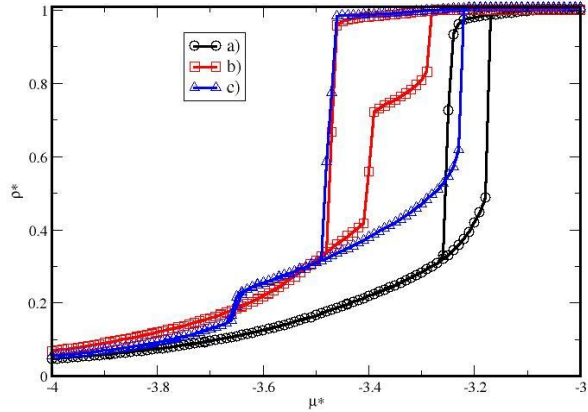

**Figure 7**. Three different isotherms obtained for complex pore structures as functions of the chemical potential, for $T^*=0.6$, $y=2.0$ and $J=0.7447$, see equation (1). The plot shown in a) corresponds to the sinusoidal structure displayed in Fig. 1(a) with radii R1=7 and R2=6 while that in b) corresponds to Fig. 1(b) with R1=7 and R2=4. The structure whose isotherms are shown in c) corresponds to that in Fig. 1(c) with R1=7 and R2=3. Circles and squares represent the adsorption and desorption processes, respectively. Lines are guides to the eyes. The axes are shown in reduced units.

As has been stressed before, the main purpose of our present work is to model and reproduce adsorption – desorption phenomena in complex structures using a simple model that has been found to be useful for other applications found in nature. Therefore we have averaged the isotherms shown in Fig. 7, as well as those listed in Table I for the three structures shown in Fig. 1 (leading to a total of 48 isotherms) to produce a single, averaged isotherm of microscopic and mesoscopic structures with multiple pore morphology and size. This average is the sum of 48 isotherms taken each $\Delta\mu^*=0.01$ from -5.5 until the condensation takes place in all the structures. The results are presented in Fig. 8, for the same conditions as the previously shown isotherms, i.e., $T^*=0.6$, $y=2.0$ and $J=0.7447$. To get $\rho^*$ we divided the number of particles present in the pores over the sum of the volumes of all the structures considered. The inset in Fig. 8, represent the radius of each structure used for the average shown in the main figure, considering all the components defined in Fig. 1 and Table I.

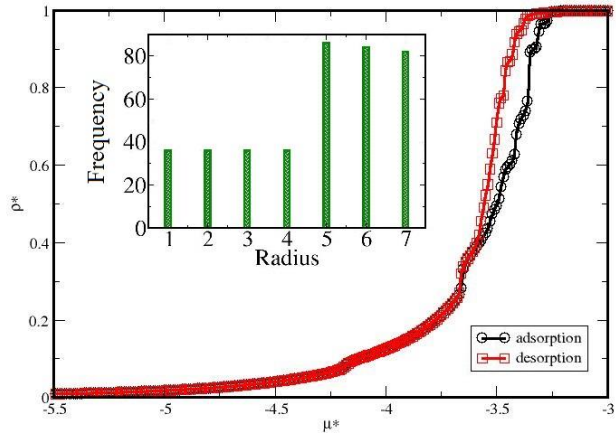

**Figure 8**. Average isotherms obtained from all those structures with radii R1 and R2 considered in this work (see Table I and Fig. 7), with $T^*$=0.6, $y$=2.0 and $J$=0.7447. Lines are guides for the eye. Axes are shown in reduced units. The inset shows the pore radii distribution of the structures used in the average. To obtain this figure we used only the radius of the cavities even if they corresponded to different geometrical structures.

As it is shown in Fig. 8, there appears a small kink at $\mu^*$=-4.2, which signals the formation of a monolayer not dissimilar to the one found in Fig. 7(b). This monolayer shows that the isotherm average must have a kink near that value of $\mu^*$. The applied relative pressure P/P⁰ jumps at $\mu^*$=-3.6, which is due to the filling of the pore with radius equal to 4σ. The following jumps, at $\mu^*$ equal to -3.5, -3.4 and -3.3 are the filling processes of the pores with radii equal to 5, 6 and 7σ respectively, since there are more of them in the average taken, as shown by the inset in Fig. 8; see also the discussion of Figs. 5 and 6. An additional feature seen in Fig. 8 is the presence of oscillations in the adsorption and desorption lines, which are found to depend on the number of isotherms used in the calculation of the average, therefore they are expected to disappear when the average is performed over a larger number of individual isotherms. However, the kinks do remain regardless of the number of isotherms used for the average, which is a feature that can be useful in the interpretation of experimentally obtained isotherms in complex pores found in nature, and their connection with their geometry.

## IV. CONCLUSIONS

In this work we report a series of computer simulations of the filling and emptying processes of complex 3D structures to understand how mesopore structures encountered in several applications and basic research problems of current interest adsorb and desorb simple gases. We studied the influence of factors such as the intensity of the surface – fluid and fluid – fluid interactions, and the geometry of the structures. The effect of increasing the surface – fluid interaction parameter *y* is the appearance of multiple layers of particles adsorbed over the surfaces, although the hysteresis cycle is not significantly affected. The structures we constructed are simple but the variations of the parameters that define them, such as the set of radii and lengths, are enough to reproduce trends found in relatively complex adsorption/desorption experiments available from the literature, and those obtained with more sophisticated simulation methodologies. To capture the essential qualitative trends found in adsorption experiments is important to carry out extensive simulations in solids with different geometries and variations of the structural and interaction parameters so that averaged isotherms can be obtained which can be sensible representations of adsorption and desorption phenomena in actual mesoporous structures. This is one of the contributions of the present work.

The approach we followed in this work was based on the modeling of relatively complex pores, made up of averages of several basic structures, while keeping the adsorbing fluid as simple as possible. Another possible avenue, which we are presently undertaking, is to start out with simple structures, such as planar - walled pores, and invest the computational power in the modeling of the adsorption of complex fluids made up, for example, of polymers, surfactants and solvent, (Gama Goicochea 2007 and 2014). However, our principal purpose here was to show how a simple model can provide useful qualitative trends that help in the understanding of experiments on adsorption in more complex systems. We expect our results to be useful for researchers working on enhanced oil recovery, in the treatment and disposing of radioactive byproducts, and in the design of new nanomaterials.

**ACKNOWLEDGEMENTS:** The authors would like to thank COMECYT for support. MABA thanks R. López – Esparza (UNISON) for discussions, and especially E. Pérez (IFUASLP) for invaluable help in the completion of this work. The use of the Aitzaloa Supercomputer cluster at UAM is acknowledged.